\documentclass[preprints,article,accept,moreauthors,pdftex]{Definitions/mdpi}

\firstpage{1} 
\pubvolume{xx}
\issuenum{1}
\articlenumber{5}
\pubyear{2020}
\copyrightyear{2020}
\history{}

\usepackage{amsmath,amsfonts,amssymb,bm}

\newcommand{\ie}{\textit{i.e.}}
\newcommand{\eg}{\textit{e.g.}}

\newcommand{\etal}{\textit{et~al.}}

\newcommand{\mB}{\mathbf{B}}

\newcommand{\mE}{\mathbf{E}}
\newcommand{\mF}{\mathbf{F}}
\newcommand{\mG}{\mathbf{G}}
\newcommand{\mH}{\mathbf{H}}

\newcommand{\mK}{\mathbf{K}}

\newcommand{\mQ}{\mathbf{Q}}

\newcommand{\mV}{\mathbf{V}}
\newcommand{\mW}{\mathbf{W}}
\newcommand{\mX}{\mathbf{X}}
\newcommand{\mY}{\mathbf{Y}}
\newcommand{\mZ}{\mathbf{Z}}

\newcommand{\vb}{\mathbf{b}}

\newcommand{\vk}{\mathbf{k}}

\newcommand{\vq}{\mathbf{q}}

\newcommand{\vv}{\mathbf{v}}

\newcommand{\valpha}{\bm \alpha}

\newcommand{\vphi}{\bm \phi}

\pdfoutput=1

\Title{
Global Structure-Aware Drum Transcription
Based on Self-Attention Mechanisms
}

\Author{
Ryoto Ishizuka$^{1}$\orcidA{}, 
Ryo Nishikimi$^{1}$\orcidB{}, 
Kazuyoshi Yoshii$^{1,2}$\orcidC{}
}

\AuthorNames{
Ryoto Ishizuka, 
Ryo Nishikimi, and 
Kazuyoshi Yoshii
}

\address{%
$^{1}$\quad
Graduate School of Informatics, Kyoto University;
\{ishizuka, nishikimi, yoshii\}@sap.ist.i.kyoto-u.ac.jp \\
$^{2}$\quad
PRESTO, Japan Science and Technology Agency (JST)
}

\abstract{
This paper describes an automatic drum transcription (ADT) method
 that directly estimates a tatum-level drum score from a music signal,
 in contrast to most conventional ADT methods
 that estimate the frame-level onset probabilities of drums.
To estimate a tatum-level score, we propose a deep transcription model
 that consists of a frame-level encoder
 for extracting the latent features from a music signal
 and a tatum-level decoder for estimating a drum score
 from the latent features pooled at the tatum level.
To capture the global repetitive structure of drum scores,
 which is difficult to learn with a recurrent neural network (RNN),
 we introduce a self-attention mechanism 
 with tatum-synchronous positional encoding into the decoder.
To mitigate the difficulty of training the self-attention-based model 
 from an insufficient amount of paired data
 and improve the musical naturalness of the estimated scores,
 we propose a regularized training method
 that uses a global structure-aware masked language (score) model 
 with a self-attention mechanism
 pretrained from an extensive collection of drum scores.
Experimental results showed that
 the proposed regularized model
 outperformed the conventional RNN-based model
 in terms of the tatum-level error rate and the frame-level F-measure,
 even when only a limited amount of paired data was available
 so that the non-regularized model underperformed the RNN-based model.
}


\begin{document}

\section{Introduction}

Automatic drum transcription (ADT)
 is one of the most important sub-tasks in automatic music transcription (AMT)
 because the drum part forms the rhythmic backbone of popular music.
In this study,
 we deal with the three main instruments of the basic drum kit:
 bass drum (BD), snare drum (SD), and hi-hats (HH).
Since these drums produce unpitched impulsive sounds,
 only the onset times are of interest in ADT.
The standard approach to ADT 
 is to estimate the activations (gains) 
 or onset probabilities of each drum from a music spectrogram \textit{at the frame level}
 and then determine the onset frames
 using an optimal path search algorithm based on some cost function~\cite{bello2005tutorial}. 
Although the ultimate goal of ADT is to estimate a human-readable symbolic drum score,
 few studies have attempted to estimate 
 the onset times of drums quantized \textit{at the tatum level}.
In this paper, 
 the ``tatum'' is defined as a tick position on the sixteenth-note-level grid 
 (four times finer than the ``beat'' on the quarter-note-level grid) 
 and the tatum times are assumed to be estimated in advance~\cite{bock2016Madmom}.

Nonnegative matrix factorization (NMF) and deep learning
 have been used for frame-level ADT~\cite{wu2018review}.
The time-frequency spectrogram of a percussive part,
 which can be separated from a music spectrogram~\cite{fabian2019open, spleeter2020},
 has a low-rank structure
 because it is composed of repeated drum sounds with varying gains.
This has motivated the use of NMF or its convolutional variants for ADT~\cite{paulus2009drum, dittmar2014real, wu2015drum, roebel2015automatic, ueda2019bayesian},
 where the basis spectra or spectrogram templates of drums are prepared
 and their frame-level activations are estimated in a semi-supervised manner.
The NMF-based approach is a physically reasonable choice,
 but the supervised DNN-based approach has recently gained much attention 
 because of its superior performance.
Comprehensive experimental comparison
 of DNN- and NMF-based ADT methods
 have been reported in~\cite{wu2018review}.
Convolutional neural networks (CNNs), for example,
 have been used for extracting local time-frequency features
 from an input spectrogram~\cite{schluter2014improved, gajhede2016convolutional, southall2017automatic, jacques2018automatic, wang2017two}.
Recurrent neural networks (RNNs) are expected to learn the temporal dynamics inherent in music
 and have successfully been used, often in combination with CNNs, 
 for estimating the smooth onset probabilities of drum sounds at the frame level~\cite{vogl2016recurrent, stables2016automatic,vogl2017drum}.
This approach, however,
 cannot learn musically meaningful drum patterns on the symbolic domain,
 and the tatum-level quantization of the estimated onset probabilities
 in an independent post-processing step
 often yields musically unnatural drum notes.

To solve this problem,
 Ishizuka \etal~\cite{ishizuka2020tatum} attempted to use the encoder-decoder architecture~\cite{sutskever2014sequence, kyunghyun2014learning}
 for frame-to-tatum ADT.
The model consisted of a CNN-based frame-level encoder 
 for extracting the latent features from a drum-part spectrogram
 and an RNN-based tatum-level decoder
 for estimating the onset probabilities of drums from the latent features
 pooled at the tatum level.
This was inspired by the end-to-end approach to automatic speech recognition (ASR),
 where the encoder acted as an acoustic model to extract the latent features from speech signals
 and the decoder acted as a language model 
 to estimate the grammatically coherent word sequences~\cite{chorowski2014end}.
Unlike ASR models, 
 the model used the temporal pooling and had no attention mechanism 
 that connected the frame-level encoder to the tatum-level decoder, 
 \ie, that aligned the frame-level acoustic features with the tatum-level drum notes,
 because the tatum times were given.
Although the tatum-level decoder was capable of learning musically meaningful drum patterns
 and favored musically natural drum notes as its output,
 the performance of ADT was limited
 by the amount of paired data of music signals and drum scores.

\begin{figure}[t]
\centerline{
\includegraphics[width=.95\linewidth]
{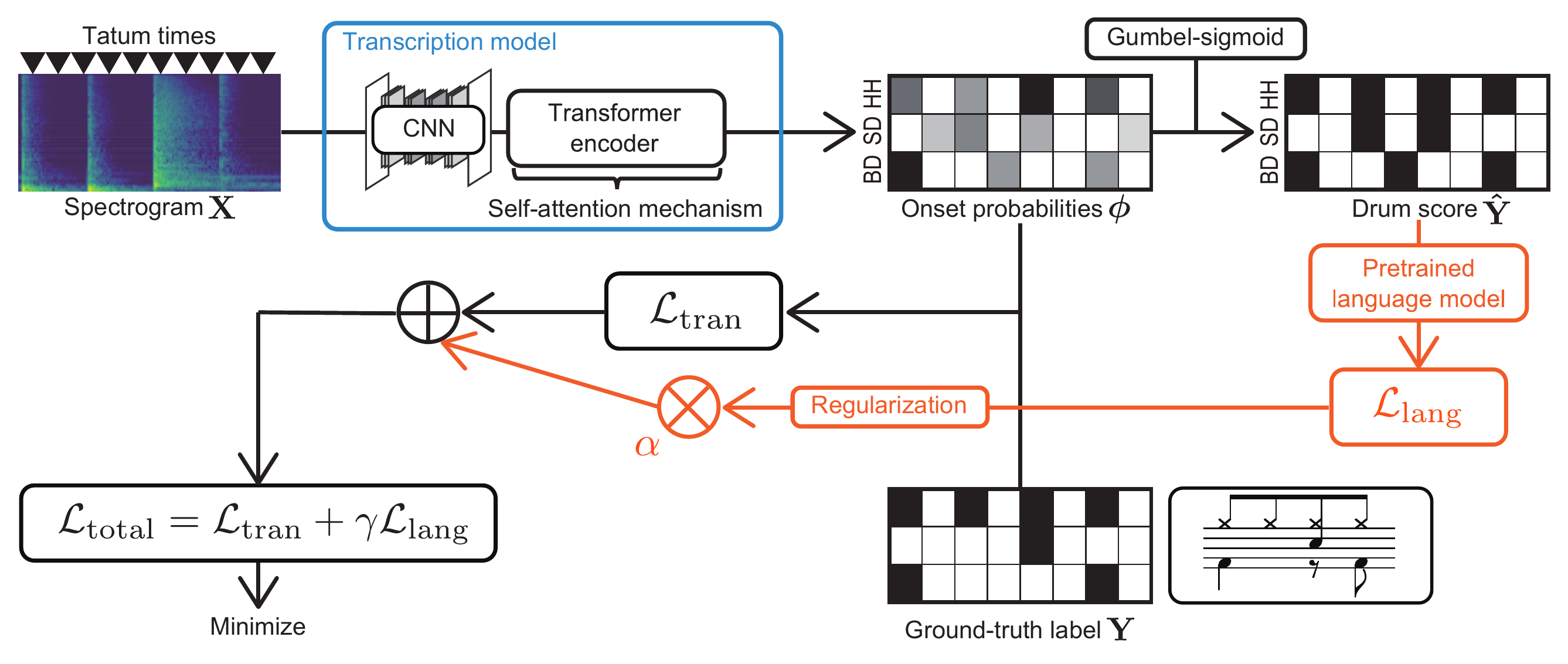}
}
\caption{
The supervised training of a neural transcription model (encoder-decoder model)
with global structure-aware regularization 
based on a pretrained language model.
}
\label{fig:model_overview}
\end{figure}

Transfer learning~\cite{bai2019learn, chen2019distilling, futami2020distilling}
 is a way of using external non-paired drum scores
 for improving the generalization capability of the encoder-decoder model.
For example,
 the encoder-decoder model could be trained in a regularized manner
 such that the output score was close to the ground-truth drum score
 and at the same time was preferred by a language model
 pretrained from an extensive collection of drum scores~\cite{ishizuka2020tatum}.
More specifically,
 a repetition-aware bi-gram model and a gated recurrent unit (GRU) model
 were used as language models
 for evaluating the probability (musical naturalness) of a drum score.
Assuming drum patterns were repeated with an interval of four beats
 as was often the case with the 4/4 time signature,
 the bi-gram model predicted the onset activations at each tatum
 by referring to those at the tatum four beats ago.
The GRU model worked better than the bi-gram model
 because it had no assumption about the time signature
 and could learn the sequential dependency of tatum-level onset activations.
Although the grammatical knowledge learned by the GRU model
 was expected to be transferred into the RNN-based decoder,
 such RNN-based models still could not learn the repetitive structure of drum patterns
 on the global time scale.
 
To overcome this limitation, in this paper
 we propose a global structure-aware frame-to-tatum ADT method
 based on an encoder-decoder model with a self-attention mechanism 
 and transfer learning (Fig.~\ref{fig:model_overview}),
 inspired by the success 
 in sequence-to-sequence tasks such as machine translation and ASR.
More specifically,
 our model involves a tatum-level decoder with a self-attention mechanism,
 where the architecture of the decoder is similar to that of the encoder of the transformer~\cite{vaswani2017attention},
 because the input and output dimensions of the decoder are the same.
To consider the temporal regularity of tatums for the self-attention computation,
 we propose a new type of positional encoding synchronized with the tatum times.
Our model is trained in a regularized manner
 such that the model output (drum score) is preferred 
 by a masked language model (MLM) with a self-attention mechanism
 that evaluates the pseudo-probability of the drum notes at each tatum
 based on both the forward and backward contexts.
We experimentally validate the effectiveness of the self-attention mechanism
 used in the decoder and/or the language model
 and that of the tatum-synchronous positional encoding.
We also investigate the computational efficiency of the proposed ADT method
 and compare it with that of the conventional RNN-based ADT method.

In Section 2 of this paper,
 we introduce related work on ADT and language modeling.
Section 3 describes the proposed method,
 and Section 4 reports the experimental results.
We conclude in Section 5 with a brief summary and mention of future work.

\section{Related Work}
This section reviews related work on ADT (Section~\ref{sec:ADT}),
 global structure-aware language models (Section~\ref{sec:global_LM}),
 and evaluation metrics for transcribed musical scores (Section~\ref{sec:evaluation_metrics}).

\subsection{Automatic Drum Transcription (ADT)}
\label{sec:ADT}

Some studies have attempted to use knowledge learned from
 an extensive collection of unpaired drum scores to improve ADT.
A language model can be trained from such data in an unsupervised manner
 and used to encourage a transcription model 
 to estimate a musically natural drum pattern.
Thompson \etal~\cite{thompson2014drum}
 used a template-based  language model for classifying audio signals
 into a limited number of drum patterns
 with a support vector machine (SVM).
Wu \etal~\cite{wu2017automatic} proposed a framework of knowledge distillation~\cite{hinton2015distilling},
 one way to achieve transfer learning~\cite{weiss2016survey}, 
 in which an NMF-based teacher model was applied
 to a DNN-based student model.

Language models have been actively used in the field of ASR.
In a classical ASR system 
 consisting of independently-trained acoustic and language models,
 the language model is used in combination 
 with the acoustic model in the decoding stage
 to generate syntactically- and semantically-natural word sequences.
The implementation of the decoder, however, is highly complicated.
In an end-to-end ASR system 
 having no clear distinction of acoustic and language models,
 only paired data can be used for training an integrated model.
Transfer learning is a promising way
 of making effective use of a language model 
 trained from huge unpaired text data~\cite{bai2019learn}.
For example, 
 a pretrained language model is used 
 for softening the target word distribution of paired speech data 
 such that not only ground-truth transcriptions
 but also their semantically-coherent variations
 are taken into account as target data
 in the supervised training~\cite{futami2020distilling}.

A frame-level language model has been used in AMT.
Raczy{\`n}ski~\etal~\cite{raczynski2013dynamic}
 used a deep belief network for modeling transitions of chord symbols
 and improved the chord recognition method based on NMF.
Sigtia \etal~\cite{sigtia2015audio} used a language model
 for estimating the most likely chord sequence
 from the chord posterior probabilities estimated by an RNN-based chord recognition system.
As pointed out in~\cite{korzeniowski2017futility,ycart2019blending}, 
 however, language models can more effectively be formulated at the tatum level
 for learning musically meaningful structures.
Korzeniowski \etal~\cite{korzeniowski2018automatic}
 used N-gram as a symbolic language model
 and improved a DNN-based chord recognition system.
Korzeniowski \etal~\cite{korzeniowski2018improved}
 used an RNN-based symbolic language model with a duration model
 based on the idea that a frame-level language model can only smooth the onset probabilities of chord symbols.
Ycart \etal~\cite{ycart2017study}
 investigated the predictive power of LSTM networks
 and demonstrated that
 a long short-term memory (LSTM) working at the level of 16th note timesteps
 could express musical structures
 such as note transitions.

A tatum-level language model has also been used in ADT.
Ueda \etal~\cite{ueda2019bayesian}
 proposed a Bayesian approach 
 using a DNN-based language model as a prior of drum scores.
Ishizuka \etal~\cite{ishizuka2020tatum}
 proposed a regularized training method
 with an RNN-based pretrained language model
 to output musically natural drum patterns.
However,
 these tatum-level language models
 cannot learn global structures,
 although the drum parts exhibit repetitive structure in music signals.

\subsection{Global Structure-Aware Language Model}
\label{sec:global_LM}

The attention mechanism is a core technology 
 for global structure-aware sequence-to-sequence learning.
In the standard encoder-decoder architecture,
 the encoder extracts latent features from an input sequence
 and the decoder recursively generates a variable number of output symbols one by one
 while referring to the whole latent features 
 with attention weights~\cite{thang2015effective, dzmitry2015neural}.
In general,
 the encoder and decoder are implemented as RNNs 
 to consider the sequential dependency of input and output symbols.
Instead, 
 the self-attention mechanism 
 that can extract global structure-aware latent features from a single sequence 
 can be incorporated into the encoder and decoder,
 leading to a non-autoregressive model called the transformer~\cite{vaswani2017attention}
 suitable for parallel computation in the training phase.

To represent the ordinal information of input symbols, 
 positional encoding vectors in addition to the input sequence
 are fed to the self-attention mechanism.
Absolute positional encoding vectors are used
 in a non-recursive sequence-to-sequence model
 based entirely on CNNs~\cite{gehring2017conv}.
Predefined trigonometric functions with different frequencies
 were proposed for representing absolute position information~\cite{vaswani2017attention}.
Sine and cosine functions are expected 
 to implicitly learn the positional relationships of symbols
 based on the hypothesis that
 there exists a linear transformation
 that arbitrarily changes the phase of the trigonometric functions.
There are some studies
 on relative position embeddings~\cite{peter2018self, zihang2019transformer, ngoc2020relative}.

Recently,
 various methods for pretraining global structure-aware language models
 have been proposed.
Embeddings from language models (ELMo)~\cite{matthew2018deep}
 is a feature-based pretraining model
 that combines forward and backward RNNs at the final layer
 to use bidirectional contexts.
However, the forward and backward inferences are separated,
 and the computation is time-consuming
 because of the recursive learning process.
Generative pretrained transformer (GPT)~\cite{radford2018improving}
 and bidirectional encoder representations from transformers (BERT)~\cite{Jacob2019BERT}
 are pretrained models based on fine tuning.
GPT is a variant of the transformer 
 trained by preventing the self-attention mechanism
 from referring to future information.
However, the inference is limited to a single direction.
BERT can jointly learn bidirectional contexts,
 and thus the masked language model (MLM) obtained by BERT is categorized as a bidirectional language model.
Because the perplexity is difficult to calculate,
 the pseudo perplexity of inferred word sequences is computed
 as described in~\cite{chen2019exploiting}.

In music generation,
 music transformer~\cite{huang2018music} uses a relative attention mechanism
 to learn long-term structure
 along with a new algorithm to reduce the memory requirement.
Pop music transformer~\cite{huang2020pop} adopts transformer-XL
 to leverage longer-range information
 along with a new data representation
 that expresses the rhythmic and harmonic structure of music.
Transformer variational autoencoder~\cite{jiang2020transformer} enables
 the joint learning of
 local representation and global structure
 based on the hierarchical modeling.
Harmony transformer~\cite{chen2019harmony} improves chord recognition
 to integrate chord segmentation with a non-autoregressive decoding method
 in the framework of musical harmony analysis.
In ADT, however,
 very few studies have focused on learning long-term dependencies,
 even though a repetitive structure can uniquely be seen in drum patterns.

\subsection{Evaluation Metrics for AMT}
\label{sec:evaluation_metrics}
Most work in AMT have conducted frame-level evaluation
 for the detected onset times of target musical instruments.
Poliner \etal~\cite{poliner2006discriminative}
 proposed comprehensive two metrics
 for frame-level piano transcription.
The first one is the accuracy rate
 defined according to Dixon's work~\cite{dixon2000computer}.
The second one is 
 the \textit{frame-level transcription error score}
 inspired by the evaluation metric
 used in multiparty speech activity detection.

Some recent studies have focused on tatum- and symbol-level evaluations.
Nishikimi \etal~\cite{nishikimi2019end} used
  the tatum-level error rate based on the Levenshtein distance
  in automatic singing transcription.
Nakamura \etal~\cite{nakamura2018towards} conducted symbol-level evaluation
 for a piano transcription system consisting 
 of multi-pitch detection and rhythm quantization.
In the multi-pitch detection stage,
 an acoustic model estimates note events
 represented by pitches, onset and offset times (in seconds), and velocities.
In the rhythm quantization stage,
 a metrical HMM with Gaussian noise (noisy metrical HMM)
 quantizes the note events on a tatum grid,
 followed by note-value and hand-part estimation.
McLeod \etal~\cite{mcleod2018evaluating}
 proposed a quantitative metric called \textit{MV2H}
 for both multipitch detection and musical analysis.
Similar to Nakamura's work,
 this metric aims to evaluate a complete musical score
 with instrument parts, a time signature and metrical structure,
 note values, and harmonic information.
The \textit{MV2H} metric is based on the principle that
 a single error should be penalized once in the evaluation phase.
In the context of ADT, in contrast,
 tatum- and symbol-level metrics have scarcely been investigated.

\section{Proposed Method}
\label{sec:proposed method}

Our goal is to estimate a drum score $\hat{\mY} \in \{0, 1\}^{M \times N}$
 from the mel spectrogram of a target musical piece
 $\mX \in {\mathbb R}_+^{F \times T}$,
 where $M$ is the number of drum instruments (BD, SD, and HH, \ie, $M {=} 3$),
 $N$ is the number of tatums,
 $F$ is the number of frequency bins,
 and $T$ is the number of time frames. 
We assume that all onset times are located on the tatum-level grid,
 and the tatum times $\mB = \{b_n\}_{n=1}^N$,
 where $1 \leq b_n \leq T$ and $b_n < b_{n+1}$,
 are estimated in advance.

In Section~\ref{sec:transcription_models},
 we explain the configuration of
 the encoder-decoder-based transcription model.
Section~\ref{sec:language_models} describes the masked language model
 as a bidirectional language model
 with the bi-gram- and GRU-based language models
 as unidirectional language models.
The regularization method is explained
 in Section~\ref{sec:regularized_training}.

\subsection{Transcription Models}
\label{sec:transcription_models}

The transcription model is used for estimating
 the tatum-level onset probabilities
 $\bm\phi \in [0, 1]^{M \times N}$,
 where $\phi_{m,n}$ represents the probability 
 that drum $m$ has an onset at tatum $n$.
The drum score $\mY$ is obtained
 by binarizing $\bm\phi$ with a threshold $\delta \in [0, 1]$.

The encoder of the transcription model is implemented with a CNN.
The mel spectrogram $\mX$ 
 is converted to latent features $\mF \in {\mathbb R}^{D_{F} \times T}$,
 where $D_{F}$ is the feature dimension.
The frame-level latent features $\mF$ are then summarized
 into tatum-level latent features $\mG \in {\mathbb R}^{D_{F} \times N}$
 through a max-pooling layer referring to the tatum times $\mB$ as follows:
\begin{align}
G_{d,n} &= \max_{\frac{b_{n-1}+b_n}{2} \le t < \frac{b_{n}+b_{n+1}}{2}} F_{d,t},
\end{align}
where $b_0 {=} b_1$ and $b_{N+1} {=} b_{N}$ are introduced for the brief expression.

\begin{figure}[t]
\centerline{
\includegraphics[width=\linewidth]
{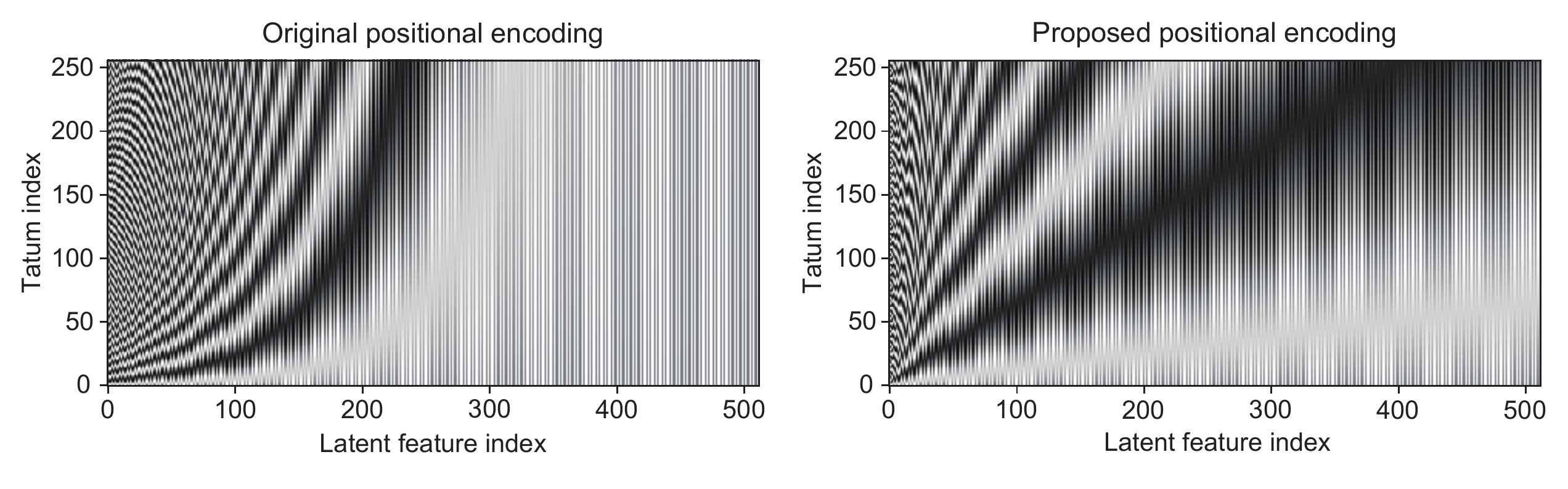}
}
\caption{
The original positional encoding
described in~\cite{vaswani2017attention} (left)
and the proposed one (right) with $D_F {=} 512$ and $N {=} 256$.
}
\label{fig:positional_encodings}
\end{figure}

The decoder of the transcription model 
 is implemented with a bidirectional GRU (BiGRU)
 or a self-attention mechanism (SelfAtt)
 followed by a fully connected layer.
The intermediate features $\mG$ are directly 
 converted to the onset probabilities at the tatum level.
In the self-attention-based decoder,
 the onset probabilities are estimated without recursive computation.
To learn the sequential dependency and global structure of drum scores,
 the positional encoding $\mE \in {\mathbb R}^{D_{F} \times N}$
 are fed into the latent features $\mG$
 to obtain extended latent features $\mZ \in {\mathbb R}^{D_{F} \times N}$.
The standard positional encodings proposed in \cite{vaswani2017attention}
 are given by
 \begin{align}
E_{d,n} &=
\begin{cases}
\sin \left( \frac{1}{10000^{2d / D_{F}}} n \right)\quad (d \equiv 0 \bmod 2), \\[4mm]
\cos \left( \frac{1}{10000^{2d / D_{F}}} n \right)\quad (d \equiv 1 \bmod 2), \\
\end{cases}
\end{align}
In this paper,
 we propose tatum-synchronous positional encodings (denoted SyncPE):
\begin{align}
E_{d,n} &=
\begin{cases}
\sin \left( \frac{\pi}{2+[ d/2 ]} n \right)\quad (d \equiv 0 \bmod 2), \\[4mm]
\cos \left( \frac{\pi}{2+[ d/2 ]} n \right)\quad (d \equiv 1 \bmod 2), \\
\end{cases}
\end{align}
where $[ \cdot ]$ represents the floor function.
As shown in Fig.~\ref{fig:positional_encodings},
 the non-linear stripes patterns appear
 in the encodings proposed in \cite{vaswani2017attention}
 because the period of the trigonometric functions increases exponentially
 with respect to the latent feature indices,
 whereas the proposed tatum-synchronous encodings
 exhibit the linear stripes patterns.

\begin{figure}[t]
\centerline{
\includegraphics[width=.95\linewidth]
{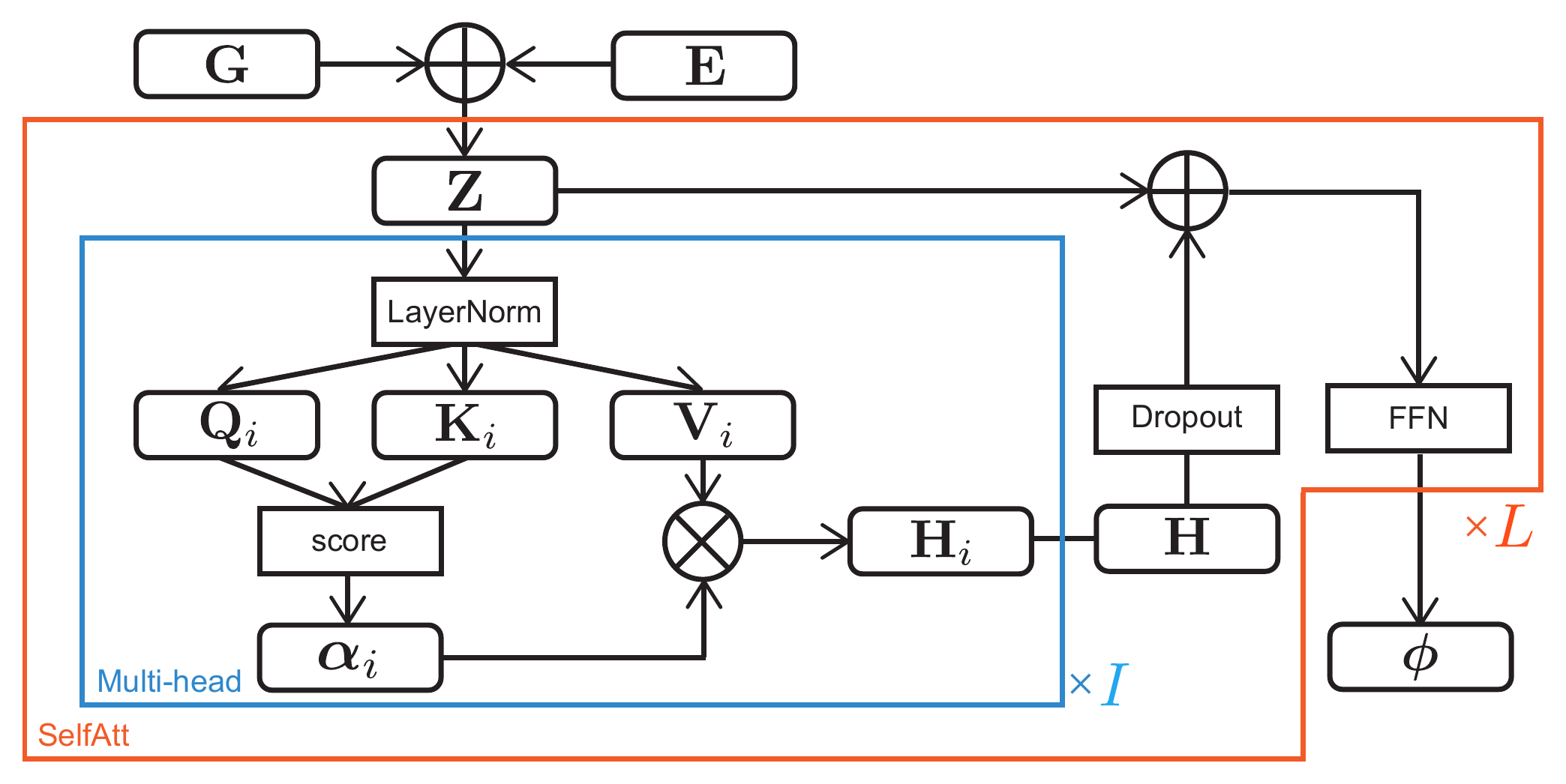}
}
\caption{
The proposed transcription model 
with a multi-head self-attention mechanism.
}
\label{fig:self-attention_decoder}
\end{figure}

As shown in Fig.~\ref{fig:self-attention_decoder},
 the extended features $\mZ$ are converted 
 to the onset probabilities $\bm\phi$
 through a stack of $L$ self-attention mechanisms with $I$ heads~\cite{vaswani2017attention}
 and the layer normalization (Pre-Norm)~\cite{ba2016layer}
 proposed for the simple and stable training of the transformer models~\cite{wang2019learning,toan2019transformer}.
For each head $i$ ($1 \leq i \leq I$),
 let $\mQ_{i} \triangleq [\vq_{i,1}, \ldots, \vq_{i,N} ] \in {\mathbb R}^{D_{K} \times N}$,
 $\mK_{i} \triangleq [\vk_{i,1}, \ldots, \vk_{i,N} ] \in {\mathbb R}^{D_{K} \times N}$, 
 and $\mV_{i} \triangleq [\vv_{i,1}, \ldots, \vv_{i,N} ] \in {\mathbb R}^{D_{K} \times N}$
 be query, key, and value matrices given by
\begin{alignat}{2}
\mQ_{i}
 &= \mW_i^{(Q)}\mathrm{LayerNorm}(\mZ) + \vb^{(Q)}_i, 
\label{equation:first_eq} \\
\mK_{i}
 &= \mW_i^{(K)}\mathrm{LayerNorm}(\mZ) + \vb^{(K)}_i, \\
\mV_{i}
 &= \mW_i^{(V)}\mathrm{LayerNorm}(\mZ) + \vb^{(V)}_i,
\end{alignat}
where
 $D_k$ is the feature dimension of each head 
 ($D_k {=} \frac{D_F}{I}$ in this paper as in~\cite{vaswani2017attention}),
 $\vq_{i} \in {\mathbb R}^{D_{K}}$,
 $\vk_{i} \in {\mathbb R}^{D_{K}}$, and
 $\vv_{i} \in {\mathbb R}^{D_{K}}$
 are query, key, and value vectors, respectively,
 $\mW^{(Q)}_i \in {\mathbb R}^{D_{K} \times D_{F}}$,
 $\mW^{(K)}_i \in {\mathbb R}^{D_{K} \times D_{F}}$, and
 $\mW^{(V)}_i \in {\mathbb R}^{D_{K} \times D_{F}}$
 are weight matrices, and 
 $\vb^{(Q)}_i \in {\mathbb R}^{D_{K} \times N}$,
 $\vb^{(K)}_i \in {\mathbb R}^{D_{K} \times N}$, and
 $\vb^{(V)}_i \in {\mathbb R}^{D_{K} \times N}$ 
 are bias vectors.
Let $\valpha \in {\mathbb R}^{N \times N}$ 
 be a self-attention matrix
 consisting of the degrees of self-relevance
 of the extended latent features $\mZ$,
 which is given by
\begin{align}
e_{i, n, n'} 
&= \frac{\vq_{i,n}^\top\vk_{i,n'}}{\sqrt{D_{K}}},
\\
\alpha_{i, n, n'} &= \frac{\exp \left(e_{i, n, n'}\right)}{\sum_{n'=1}^{N}\exp \left( e_{i, n, n'} \right)},
\end{align}
where
 $^\top$ represents the matrix or vector transpose 
 and $n$ and $n'$ represent the feature indices of $\mQ$ and $\mK$, respectively.
Let $\mH \triangleq [\mH_1, \ldots, \mH_I]\in\mathbb{R}^{D_F \times N}$
 be a feature matrix obtained by concatenating all the heads, 
 where $\mH_i \triangleq \mV_i{\valpha}_i^\top\in\mathbb{R}^{D_K \times N}$.

The extended latent features and the extracted features $\mH$
 with Dropout ($p = 0.1$)~\cite{srivastava2014dropout}
 are then fed into a feed forward network (FFN)
 with a rectified linear unit (ReLU) as follows:
\begin{align}
 \mZ
 \leftarrow 
 \mW_2^{(H)}{\max}\left\{0, \mW_1^{(H)}\left(\mathrm{Dropout}\left(\mH\right) + \mZ \right) + \vb_1^{(H)}\right\} + \vb_2^{(H)}, \label{equation:last_eq}
\end{align}
where
 $\mW_1^{(H)} \in {\mathbb R}^{D_{\mathrm{FFN}} \times D_F}$ and
 $\mW_2^{(H)} \in {\mathbb R}^{D_{F} \times D_{\mathrm{FFN}}}$
 are weight matrices,
 $\vb_1^{(H)} \in {\mathbb R}^{D_{\mathrm{FFN}} \times N}$ and $\vb_2^{(H)} \in {\mathbb R}^{D_F \times N}$
 are bias vectors,
 and $D_{\mathrm{FFN}}$ is the dimension of the output.
Eq.~(\ref{equation:first_eq}) to Eq.~(\ref{equation:last_eq}) 
 are repeated $L$ times with different parameters.
The onset probabilities $\bm\phi$ are finally calculated as follows:
\begin{align}
\vphi &= \sigma \left( \mW_3^{(H)}\mZ + \vb_3^{(H)} \right),
\end{align}
where
 $\sigma(\cdot)$ is a sigmoid function,
 $\mW_3^{(H)} \in {\mathbb R}^{M \times D_{F}}$ is a weight matrix, and
 $\vb_3^{(H)} \in {\mathbb R}^{M \times N}$ is a bias vector.

\subsection{Language Models}
\label{sec:language_models}
\begin{figure}[t]
\centerline{
\includegraphics[width=\linewidth]
{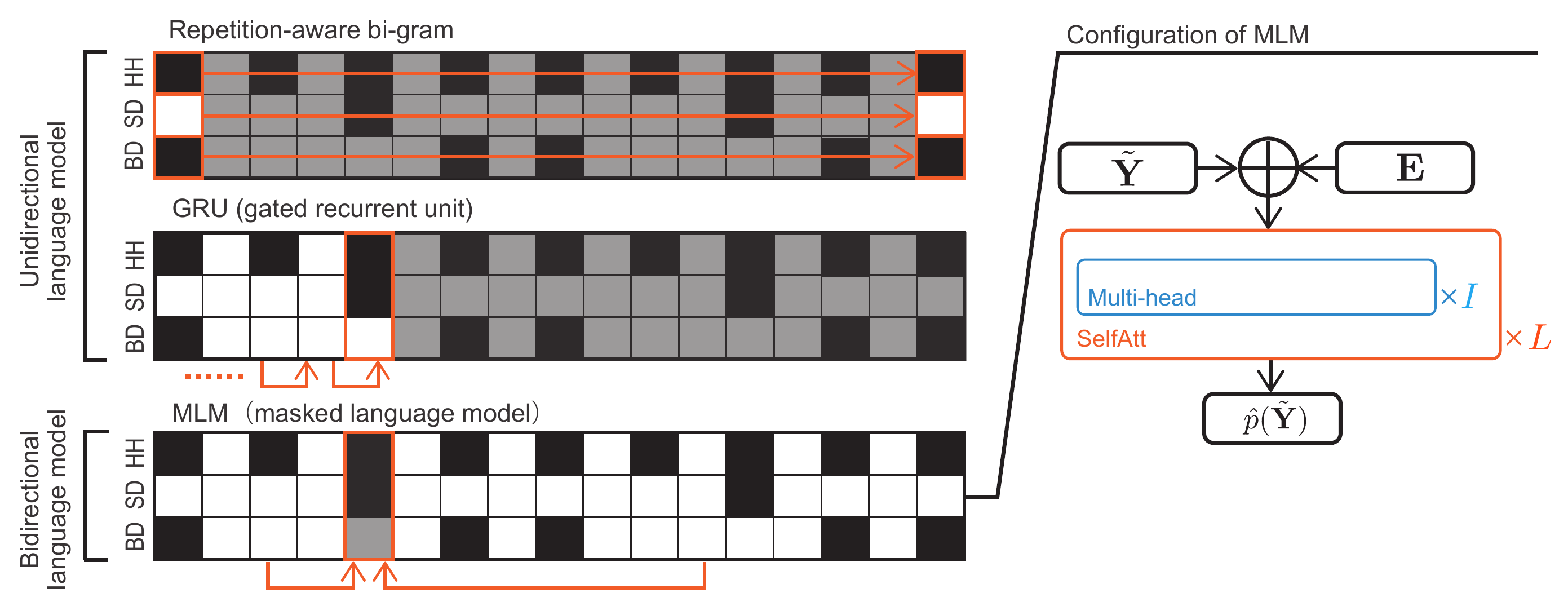}
}
\caption{
Language models used for regularized training of a transcription model.
}
\label{fig:LMs}
\end{figure}

The language model is used
 for estimating the generative probability (musical naturalness)
 of an arbitrary existing drum score $\tilde\mY$\footnote{
 For brevity, we assume that only one drum score is used as training data.
 In practice, a sufficient amount of drum scores are used.
 }.
In this study,
 we use unidirectional language models
 such as the repetition-aware bi-gram model and GRU-based model
 proposed in~\cite{ishizuka2020tatum}
 and a masked language model (MLM),
 a bidirectional language model
 proposed for pretraining in BERT~\cite{Jacob2019BERT} (Fig.~\ref{fig:LMs}).

The unidirectional language model
 is trained beforehand in an unsupervised manner
 such that the following negative log-likelihood
 ${\mathcal L}_{\mathrm{lang}}^{\mathrm{uni}} (\tilde{\mY})$
 is minimized:
\begin{align}
\mathcal{L}_{\mathrm{lang}}^{\mathrm{uni}} (\tilde{\mY}) 
 = - \log p (\tilde{\mY})
 = - \sum_{n=1}^{N} \log p (\tilde{Y}_{:,n} | \tilde{Y}_{:,1:n-1}),
\label{eq:l_lang_uni}
\end{align}
where ``$i{:}j$'' represents a set of indices from $i$ to $j$
 and ``$:$'' represents all possible indices.
In the repetition-aware bi-gram model (top figure in Fig.~\ref{fig:LMs}) 
 assuming that target musical pieces have the 4/4 time signature,
 the repetitive structure of a drum score is formulated as follows:
\begin{align}
 p(\tilde{Y}_{:,n} | \tilde{Y}_{:,1:n-1})
 =
 \prod_{m=1}^M p(\tilde{Y}_{m,n} | \tilde{Y}_{m,n-16})
 =
 \prod_{m=1}^M \pi_{\tilde{Y}_{m,n-16},\tilde{Y}_{m,n}},
\end{align}
where $\pi_{A,B} \ (A, B \in \{0,1\})$ represents the transition probability from $A$ to $B$.
Note that this model assumes the independence of the $M$ drums.
In the GRU model (middle figure in Fig.~\ref{fig:LMs}),
 $p(\tilde{Y}_{:,n} | \tilde{Y}_{:,1:n-1})$ is directly calculated
 using an RNN.

The MLM is capable of learning the global structure of drum scores
 (bottom figure in Fig. \ref{fig:LMs}).
In the training phase,
 drum activations at randomly selected 15$\%$ of tatums in $\tilde\mY$ are masked
 and the MLM is trained such that those activations are predicted as accurately as possible. 
 The loss function $\mathcal{L}_{\mathrm{lang}}^{\mathrm{bi}}(\tilde\mY)$ to be minimized 
 is given by
\begin{align}
\hat{p}(\tilde \mY_n) &= p(\tilde Y_{:, n} | \tilde Y_{:,1:n-1}, \tilde Y_{:, n+1:N}), \\
\mathcal{L}_{\mathrm{lang}}^{\mathrm{bi}}(\tilde\mY)
&= - \sum_{n=1}^N \log \hat{p}(\tilde \mY_n).
\label{eq:l_lang_bi}
\end{align}

\subsection{Regularized Training}
\label{sec:regularized_training}

To consider the musical naturalness of the estimated score $\hat{\mY}$
 obtained by binarizing $\bm\phi$,
 we use the language model-based regularized training method~\cite{ishizuka2020tatum}
 that minimizes
\begin{align}
{\mathcal L}_{\mathrm{total}}
&=
{\mathcal L}_{\mathrm{tran}}(\bm\phi|\mY)
+ 
\gamma {\mathcal L}_{\mathrm{lang}}^{*}(\hat\mY),
\label{sec:l_total}
\end{align}
where 
 $\hat\mY$ is a ground-truth score,
 $\gamma > 0$ is a weighting factor,
 the symbol * denotes ``uni`` or ``bi``,
 and ${\mathcal L}_\mathrm{tran} (\bm\phi|\hat\mY)$ is 
 the modified negative log-likelihood given by
\begin{align}
\mathcal{L}_{\mathrm{tran}}(\bm\phi|\mY)
&=
- \sum_{m=1}^{M} \sum_{n=1}^{N}
\left \{\beta_m Y_{m,n}\log \phi_{m,n} 
+ 
(1  -  Y_{m,n}) \log (1  -  \phi_{m,n}) \right \},
\label{sec:l_tran}
\end{align}
where
 $\beta_m > 0$ is a weighting factor 
 compensating for the imbalance of the numbers of onset and non-onset tatums.

To use backpropagation for optimizing the transcription model,
 the binary score $\mY$ should be obtained 
 from the soft representation $\bm\phi$ in a differentiable manner
 instead of simply binarizing $\bm\phi$ with a threshold.
We thus use a differentiable sampler
 called the Gumbel-sigmoid trick~\cite{tsai2018learning}, as follows:
\begin{align}
    \eta_{m,n}^{(k)} &\sim
    {\rm Uniform}(0,1),
    \\
    \psi_{m,n}^{(k)} &=
    -\log \left\{
    -\log\left( \eta_{m,n}^{(k)} \right)
    \right\},
    \\
    \hat{Y}_{m,n} &=
    \sigma \left\{
    \frac{\phi_{m,n}+\psi_{m,n}^{(1)}-\psi_{m,n}^{(2)}}{\tau}
    \right\},
\end{align}
where
 $k {=} 1, 2,$ and $\tau > 0$ is a temperature ($\tau {=} 0.2$ in this paper).
Note that the pretrained language model
 is used as a fixed regularizer in the training phase
 and is not used in the prediction phase.

\begin{figure}[t]
\centerline{
\includegraphics[width=\linewidth]
{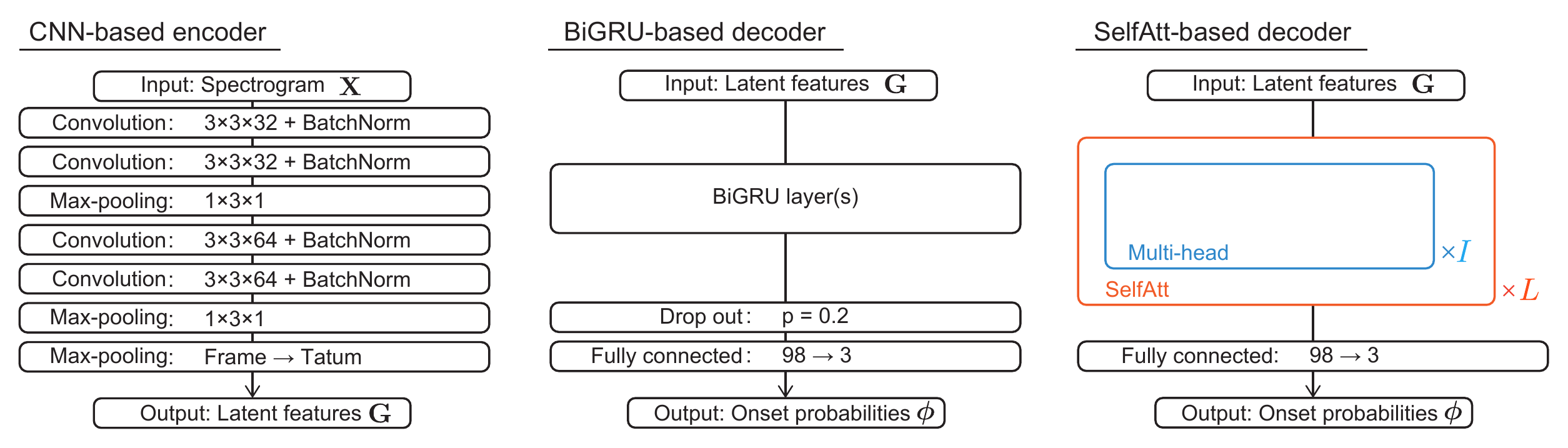}
}
\caption{
Configurations of the transcription models (CNN-BiGRU and CNN-SelfAtt).
}
\label{fig:encoder_decoder}
\end{figure}

\section{Evaluation}

This section reports the comparative experiments
 conducted for evaluating the proposed ADT method
 and investigates the effectiveness of the self-attention mechanism
 and that of the MLM-based regularized training.

\subsection{Evaluation Data}

We used the Slakh2100-split2 (Slakh)~\cite{manilow2019cutting}
 and the RWC Popular Music Database (RWC)~\cite{goto2002rwc} for evaluation
 because these datasets include ground-truth beat times.
The Slakh dataset
 contains 2100 musical pieces
 whose audio signals were 
 synthesized from the Lakh MIDI dataset~\cite{colin2019learning}
 using professional-grade virtual instruments,
 and the RWC dataset contains 100 Japanese popular songs.
All music signals were sampled at 44.1 kHz.
The onset times of BD, SD, and HH ($M=3$) were extracted 
 as ground-truth data
 from the synchronized MIDI files provided for these datasets.
To make ground-truth drum scores,
 each onset time was quantized to the closest tatum time
 (the justification of this approach is discussed 
 in Section~\ref{sec:justification}).
Only musical pieces whose drum onset times and tatum times 
 had been annotated correctly as ground-truth data
 were used for evaluation.

As to the Slakh dataset,
 we used 2010 pieces, which were split 
 into 1449, 358, and 203 pieces
 as training, validation, and test data, respectively.
As to the RWC dataset,
 we used 65 songs for 10-fold cross validation,
 where 15$\%$ of training data 
 was used as validation data in each fold.
Since we here aim to validate the effectiveness of the language model-based regularized training
 for self-attention-based transcription models on the same dataset (Slakh or RWC),
 investigation of the cross-corpus generalization capability (portability) 
 of language models is beyond the scope of the paper and left as future work.

For each music signal,
 a drum signal was separated with Spleeter~\cite{spleeter2020}
 and the tatum times were estimated with madmom~\cite{bock2016Madmom}
 or given as oracle data.
The spectrogram of a music or drum signal
 was obtained using short-time Fourier transform (STFT)
 with a Hann window of 2048 points (46 ms)
 and a shifting interval of 441 points (10 ms).
We used mel-spectrograms as input features
 because they have successfully been used 
 for onset detection~\cite{schluter2014improved}
 and CNN-based ADT~\cite{jacques2018automatic}.
The mel-spectrogram was computed using a mel-filter bank 
 with 80 bands from 20 Hz to 20,000 Hz
 and normalized so that the maximum volume was 0 db.
A stack of music and drum mel-spectrograms was fed 
 into a transcription model.
 
\subsection{Model Configurations}
\label{sec:model_configuration} 
The configurations of the two transcription models 
 (CNN-BiGRU and CNN-SelfAtt(-SyncPE) described in Section~\ref{sec:transcription_models})
 are shown in Figure~\ref{fig:encoder_decoder} 
 and Table~\ref{table:hyperparameters_tran}.
The encoders were the same CNN consisting of four convolutional layers
 with a kernel size of $3\times3$ 
 and the decoders were based on the BiGRU and the multi-head self-attention mechanism.
The influential hyperparameters
 were automatically determined with a Bayesian optimizer 
 called Optuna~\cite{optuna_2019}
 for the validation data in the Slakh dataset
 or with 10-fold cross validation in the RWC dataset
 under a condition that $D_{\mathrm{FFN}} = 4 D_{\mathrm{F}}$.
As a result, CNN-SelfAtt
 had about twice as many parameters as CNN-BiGRU.
In the training phase, the batch size was set to 10,
 and the max length was set to 256 for CNN-SelfAtt.
We used the AdamW optimizer~\cite{Loshchilov2019decoupled}
 with an initial learning rate of $10^{-3}$.
The learning rate of CNN-SelfAtt
 was changed according to~\cite{vaswani2017attention},
 and $warmup\_steps$ was set to 4000.
To prevent over-fitting,
 we used weight regularization ($\lambda = 10^{-4}$),
 drop-out just before all the fully connected layers of CNN-BiGRU ($p = 0.2$)
 and each layer of CNN-SelfAtt ($p = 0.1$),
 and tatum-level SpecAugment~\cite{du2019data} for the RWC dataset,
 where 15\% of all tatums were masked in the training phase.
The weights of the convolutional and BiGRU layers were initialized
 based on~\cite{kaiming2015delving},
 the fully connected layer was initialized by the sampling from ${\rm Uniform}(0,1)$,
 and the biases were initialized to 0.
In the testing phase,
 the average of the ten parameters
 before and after the epoch
 that achieved the smallest loss for the validation data
 was used in CNN-SelfAtt.
The threshold for $\bm\phi$ was set to $\delta = 0.2$.

The configurations of the three language models 
 (bi-gram, GRU, and MLM(-SyncPE) described in Section~\ref{sec:language_models})
 are shown in Table~\ref{table:hyperparameters_lang}.
Each model was trained with 512 external drum scores 
 of Japanese popular songs and Beatles songs.
To investigate the impact of the data size for predictive performance,
 each model was also trained by using only randomly selected 51 scores.
The influential hyperparameters of the neural language models,
 \ie, the number of layers and the dimension of hidden states in the GRU model 
 and $h$, $l$, $D_{\mathrm{F}}$, and $D_{\mathrm{FFN}}$ in the MLM,
 were automatically determined with Optuna~\cite{optuna_2019}
 via 3-fold cross validation with the 512 scores
 under a condition that $D_{\mathrm{FFN}} = 4 D_{\mathrm{F}}$.
As a result, the MLM had about twice as many parameters as the GRU model.
The bi-gram model was defined by only two parameters $\pi_{0,1}$ and $\pi_{1,1}$.

\begin{table}[t]
\centering
\caption{
Hyperparameters of the decoders of the two transcription models optimized by Optuna.
}
\begin{tabular}{c|cc|c|cccc|c} \toprule
& \multicolumn{3}{c|}{BiGRU} & \multicolumn{5}{c}{SelfAtt}\\ 
& Layer & Dim & Size & $h$ & $l$ & $D_{F}$ & $D_{\mathrm{FFN}}$ & Size \\\midrule
Slakh &  1 & 131 & 573k & 2 & 8 & 96 & 384 & 1.03M \\
RWC & 3 & 98 & 774k & 8 & 7 & 120 & 480 & 1.38M \\
\bottomrule
\end{tabular}
\label{table:hyperparameters_tran}

\bigskip

\centering
\caption{
Hyperparameters of the three language models optimized by Optuna.
}
\begin{tabular}{c|cc|c|cccc|c} \toprule
Bi-gram & \multicolumn{3}{c|}{GRU} & \multicolumn{5}{c}{MLM-SyncPE}\\ 
Size & Layer & Dim & Size & $h$ & $l$ & $D_{F}$ & $D_{\mathrm{FFN}}$ & Size \\\midrule
2 & 3 & 64 & 63k & 4 & 8 & 112 & 448 & 1.25M \\
\bottomrule
\end{tabular}
\label{table:hyperparameters_lang}
\end{table}

\begin{figure}[t]
\centerline{
\includegraphics[width=\linewidth]
{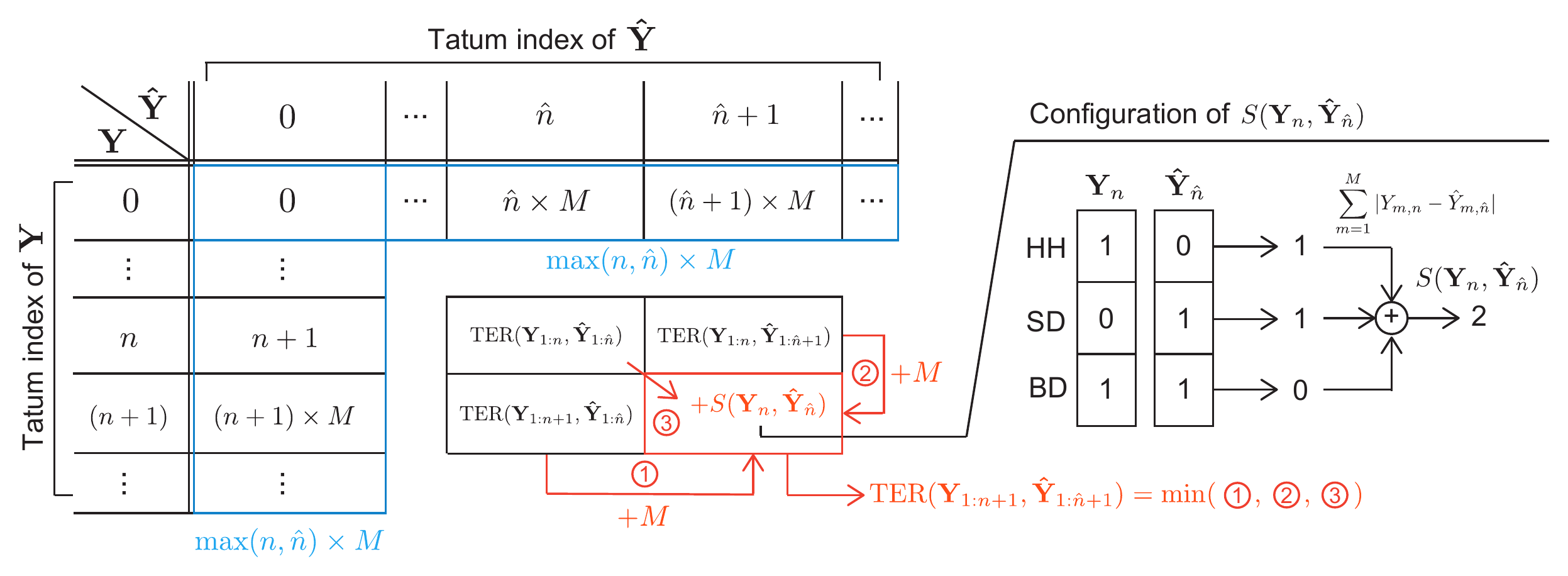}
}
\caption{
Computation of tatum-level error rate (TER)
based on dynamic programming.
}
\label{fig:TER}
\end{figure}
\begin{figure}[t]
\centerline{
\includegraphics[width=\linewidth]
{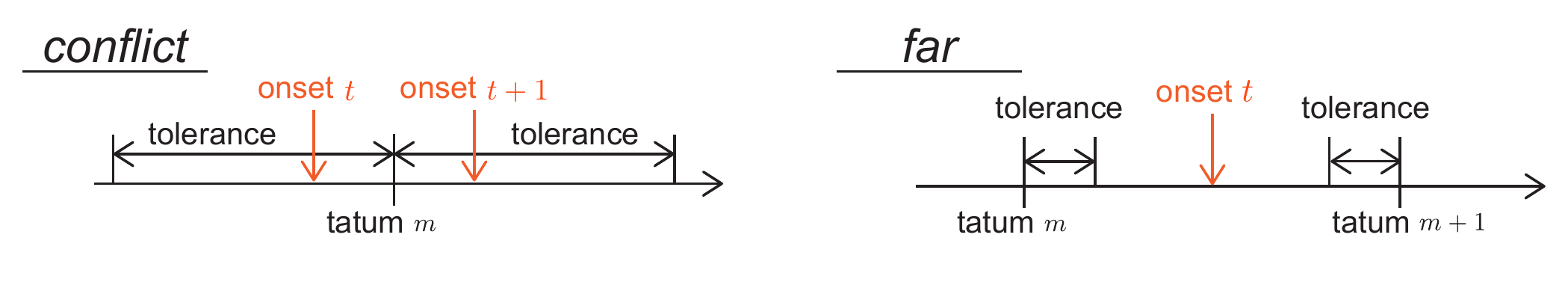}
}
\vspace{-4mm}
\caption{
Two groups of undetectable onset times.
}
\label{fig:impossible}
\end{figure}

\subsection{Evaluation Metrics}

The performance of ADT was evaluated 
 at the frame and tatum levels.
The frame-level F-measure ($\mathcal{F}$) 
 is defined as the harmonic mean
 of the precision rate $\mathcal{P}$ and the recall rate $\mathcal{R}$:
\begin{align}
  \mathcal{P} = \frac{\mathrm{N_C}}{\mathrm{N_E}},\quad
  \mathcal{R} = \frac{\mathrm{N_C}}{\mathrm{N_G}},\quad
  \mathcal{F} = \frac{2\mathcal{R}\mathcal{P}}{\mathcal{R} + \mathcal{P}}, \label{equation:F-measure}
\end{align}
where $N_\mathrm{E}$, $N_\mathrm{G}$, and $N_\mathrm{C}$
 are the number of estimated onset times,
 that of ground-truth onset times, 
 and that of correctly-estimated onset times,
 respectively,
 and the error tolerance was set to 50 ms.
Note that $\mathcal{F} = 100\%$ means perfect transcription.
For the tatum-level evaluation,
 we propose a tatum-level error rate (TER)
 based on the Levenshtein distance.
Note that all the estimated drum scores were concatenated
 and then the frame-level F-measure and TER were computed for the whole dataset.
As shown in Fig.~\ref{fig:TER},
 the TER between 
 a ground-truth score 
 $\mY \triangleq \mY_{1:N} \in \{0, 1\}^{M \times N}$ 
 with $N$ tatums
 and an estimated score 
 $\hat{\mY} \triangleq \hat{\mY}_{1:\hat{N}} \in \{0, 1\}^{M \times \hat{N}}$ with $\hat{N}$ tatums,
 denoted by $\mathrm{TER}(\mY_{1:N}, \hat{\mY}_{1:\hat{N}})$,
 is computed via dynamic programming 
 as follows:
\begin{alignat}{2}
\mathrm{TER}(\mY_{1:n+1}, \hat{\mY}_{1:\hat{n}+1}) 
&=
\begin{cases}
\max (n, \hat{n}) \times M \quad &(n=0\ \mathrm{or}\ \hat{n}=0),
\\[1mm]
\min
\begin{cases}
\mathrm{TER}(\mY_{1:n}, \hat{\mY}_{1:\hat{n}+1}) + M \\[1mm]
\mathrm{TER}(\mY_{1:n+1}, \hat{\mY}_{1:\hat{n}}) + M \\[1mm]
\mathrm{TER}(\mY_{1:n}, \hat{\mY}_{1:\hat{n}}) 
+ S(\mY_{n}, \hat{\mY}_{\hat{n}}) \\[1mm]
\end{cases}
\quad &(\mathrm{otherwise}),
\end{cases}\\
S(\mY_{n}, \hat{\mY}_{\hat{n}})
&= 
\sum_{m=1}^M | Y_{m, n} - \hat{Y}_{m, \hat{n}} |,
\end{alignat}
where $S(\mY_{n}, \hat{\mY}_{\hat{n}})$ 
 represents the sum of the Manhattan distances
 between the ground-truth activations $\mY_{n}$ at tatum $n$ 
 and the estimated activations $\hat{\mY}_{\hat{n}}$ at tatum $\hat{n}$.
Note that $\hat{N}$ might be different from $N$
 when the tatum times were estimated with madmom
 and that $\mathrm{TER}(\mY_{1:N}, \hat{\mY}_{1:\hat{N}}) = 0$
 does not mean perfect transcription 
 as discussed in Section~\ref{sec:justification}.
The comprehensive note-level evaluation measures were proposed 
 for AMT~\cite{nakamura2018towards, mcleod2018evaluating},
 but were not used in our experiment
 because ADT focuses on only the onset times of note events.

\subsection{Justification of Tatum-Level Drum Transcription}
\label{sec:justification}

We validated the appropriateness of tatum-level ADT
 because some kinds of actual onset times cannot be detected in principle
 under an assumption that the onset times of each drum
 are exclusively located on the sixteenth-note-level grid.
As shown in Fig.~\ref{fig:impossible},
 such undetectable onsets are (doubly) categorized into two groups:
 \textit{conflict} and \textit{far}.
If multiple onset times are close to the same tatum time,
 only one onset time can be detected,
 \ie, the other onset times are undetectable
 and categorized into the \textit{conflict} group.
Onset times that are not within 50 ms from the closest tatum times
 are categorized into the \textit{far} group.
In the tatum-level ADT,
 the onset times of these groups remain undetected 
 even when $\mathrm{TER}(\mY_{1:N}, \hat{\mY}_{1:\hat{N}}) = 0$.
 
Table~\ref{table:statistical_features} shows
 the ratio of undetectable onset times in each group
 to the total number of actual onset times
 when the estimated or ground-truth beat times were used for quantization.
The total ratio of undetectable onset times was sufficiently low.
This justifies the sixteenth-note-level quantization of onset times,
 at least for the majority of typical popular songs used in our experiment.
Note that our model cannot deal with triplet notes.

Since the tatum times are assumed to be given,
 we evaluated the beat tracking performance of madmom~\cite{bock2016Madmom} in terms of the F-measure in the same way as Eq.~(\ref{equation:F-measure}).
The mir\_eval library was used 
 for computing $\mathcal{P}$, $\mathcal{R}$, and $\mathcal{F}$.
The F-measure for the 203 pieces of the test data 
 in the Slakh dataset was 92.5\% 
 and that for the 65 songs in the RWC dataset was 96.4\%. 
 
\begin{table}[t]
\centering
\caption{
Ratios of undetectable onset times.
}
\vspace{-1mm}
\begin{tabular}{c|ccc|ccc} \toprule
\multirow{2}{*}{Dataset} &\multicolumn{3}{c|}{madmom} & 
\multicolumn{3}{c}{Ground-truth}
\\
& \textit{conflict} & \textit{far} & \textit{\textit{conflict} $\cup$ \textit{far}} &
\textit{conflict} & \textit{far} & \textit{\textit{conflict} $\cup$ \textit{far}}
\\ \midrule
Slakh & 0.86\% & 0.16\% & 1.02\% & 1.30\% & 0.35\% & 1.64\% \\ 
RWC & 0.43\% & 0.23\% & 0.65\% & 1.19\% & 0.29\% & 1.48\% \\ 
\bottomrule
\end{tabular}
\label{table:statistical_features}
\end{table}
\begin{table}[t]
\centering
\caption{
Perplexities obtained by the language models.
``MLM`` and  ``MLM-SyncPE``
 are MLMs with the conventional and proposed tatum-synchronous positional encodings, respectively.}
\begin{tabular}{c|cccc|cccc} \toprule
Language model & \multicolumn{2}{c}{Bi-gram} & \multicolumn{2}{c|}{GRU} & \multicolumn{2}{c}{MLM} & \multicolumn{2}{c}{MLM-SyncPE}\\
Dataset size & 51 & 512 & 51 & 512 & 51 & 512 & 51 & 512 \\ \midrule
Slakh 
& 1.278 & 1.265 & 1.357 & 1.170 & 1.180 & 1.050 & 1.124 & 1.049 \\ 
RWC
& 1.374 & 1.369 & 1.473 & 1.273 & 1.289 & 1.086 & 1.217 & 1.085 \\
\bottomrule
\end{tabular}
\label{table:PPLs}
\end{table}

\begin{table}[t]
\centering
\setlength{\tabcolsep}{1.4mm}
\caption{
Weighting factors optimized by Optuna.
$m {=} 0$, $1$, and $2$ represent BD, SD, and HH, respectively. 
}
\begin{tabular}{c|ccccccccc|cccccc} \toprule
&\multicolumn{9}{c|}{CNN-BiGRU} & \multicolumn{6}{c}{CNN-SelfAtt}\\ 
&\multirow{2}{*}{$\beta^{\star}_0$} & \multirow{2}{*}{$\beta^{\star}_1$} & \multirow{2}{*}{$\beta^{\star}_2$} & \multirow{2}{*}{$\beta_0$} & \multirow{2}{*}{$\beta_1$} & \multirow{2}{*}{$\beta_2$} &
Bi-gram & GRU & MLM
& \multirow{2}{*}{$\beta_0$}& \multirow{2}{*}{$\beta_1$} & \multirow{2}{*}{$\beta_2$} &
Bi-gram & GRU & MLM \\
&&&&&&& \multicolumn{3}{c|}{$\gamma$} &
&&& \multicolumn{3}{c}{$\gamma$} \\ \midrule
Slakh & 0.67 & 2.00 & 1.77 & 1.07 & 0.19 & 0.40 & 0.10 & 0.05 & 0.01 &
      0.62 & 0.92 & 0.90 & 1.02 & 0.07 & 1.25 \\
RWC   & 6.22 & 8.09 & 6.48 & 0.50 & 0.32 & 0.71 & 0.05 & 0.04 & 0.26 &
      0.69 & 0.92 & 0.55 & 1.10 & 0.05 & 1.31 \\
\bottomrule
\end{tabular}
\label{table:weighting_factors}
\end{table}

\begin{table*}[t]
\centering
\setlength{\tabcolsep}{1.5mm}
\caption{
Training times per song (s/song),
 F-measures ($\%$),
 and TERs obtained by the conventional and proposed methods.
}
\begin{tabular}{c|ll|r|ccccc|ccccc} \toprule
& \multicolumn{2}{l|}{Decoder} & \multirow{2}{*}{Time} & \multicolumn{5}{c|}{madmom} & \multicolumn{5}{c}{Ground-truth} \\
& \multicolumn{2}{l|}{~~+ Language model} & & BD & SD & HH & Total & TER & BD & SD & HH & Total & TER \\\midrule
\multirow{7}{*}{Slakh} & BiGRU~\cite{vogl2017drum} & & 55.8
& 93.6 & 92.7 & 71.5 & 85.9 & 20.6 & 93.0 & 92.5 & 71.4 & 85.6 & 8.4 \\ 
&
\multicolumn{2}{l|}{BiGRU} & 15.3
& 95.6 & 90.2 & 75.5 & 87.1 & 19.6 & 95.2 & 90.2 & 75.8 & 87.1 & 7.2 \\
&
~~+ MLM-SyncPE & (512) & 137.0
& 95.3 & 90.7 & 78.4 & 88.1 & 19.0 & 94.7 & 90.9 & 77.6 & 87.8 & 7.1 \\
& \multicolumn{2}{l|}{SelfAtt-SyncPE} & 26.7
& 95.8 & 93.1 & 79.9 & 89.6 & 18.7 & 95.6 & 92.9 & 79.5 & 89.3 & 6.8 \\
&
~~+ Bi-gram & (512) & 15.6
& 95.8 & 92.7 & 80.5 & 89.7 & {\bf 18.5} & 95.4 & 93.1 & 80.5 & 89.7 & {\bf 6.3} \\
&
~~+ GRU & (512) & 15.7
& 96.4 & 93.0 & 80.5 & {\bf 90.0} & {\bf 18.5} & 96.1 & 92.9 & 80.4 & 89.8 & 6.4 \\
&
~~+ MLM-SyncPE & (512) & 42.0
& 96.1 & 93.2 & 80.7 & {\bf 90.0} & {\bf 18.5} & 95.8 & 93.3 & 80.9 & {\bf 90.0} & {\bf 6.3} \\ 
\midrule
\multirow{7}{*}{RWC} & BiGRU~\cite{vogl2017drum} & & 58.2
& 86.0 & 74.0 & 70.5 & 76.8 & 16.7 & 86.6 & 74.1 & 70.9 & 77.2 & 12.3 \\ 
& \multicolumn{2}{l|}{BiGRU} & 24.5
& 86.9 & 76.2 & 77.9 & 80.3 & 15.2 & 86.7 & 76.5 & 76.9 & 80.0 & 11.4 \\
&
~~+ MLM-SyncPE & (512) & 130.1
& 88.0 & 76.5 & 79.7 & {\bf 81.4} & {\bf 14.0} & 88.1 & 76.5 & 79.3 & {\bf 81.3} & {\bf 10.3} \\
& \multicolumn{2}{l|}{SelfAtt-SyncPE} & 30.5
& 87.5 & 76.4 & 72.6 & 78.8 & 17.0 & 88.0 & 75.6 & 72.9 & 78.8 & 13.2 \\
&
~~+ Bi-gram & (512) & 31.8
& 86.0 & 76.5 & 69.6 & 77.4 & 16.8 & 87.0 & 76.0 & 69.6 & 77.5 & 12.7 \\
&
~~+ GRU & (512) & 24.1
& 87.6 & 76.2 & 73.2 & 79.0 & 16.6 & 87.8 & 76.3 & 73.9 & 79.4 & 12.5 \\
&
~~+ MLM-SyncPE & (512) & 51.6
& 88.1 & 74.9 & 75.4 & 79.5 & 16.2 & 87.9 & 74.3 & 71.7 & 78.0 & 12.4 \\ 
\bottomrule
\end{tabular}
\label{table:proposed_method_measures}
\vspace{-2mm}
\end{table*}

\begin{table*}[t]
\centering
\caption{
F-measures ($\%$) and TERs obtained 
 by CNN-SelfAtt with the conventional positional encodings
 and CNN-SelfAtt-SyncPE with the proposed tatum-synchronous positional encodings.
}
\begin{tabular}{c|ll|ccccc|ccccc} \toprule
& \multicolumn{2}{l|}{\multirow{2}{*}{Decoder}} & \multicolumn{5}{c|}{madmom} & \multicolumn{5}{c}{Ground-truth} \\
& & & BD & SD & HH & Total & TER & BD & SD & HH & Total & TER \\\midrule
\multirow{2}{*}{Slakh} & \multicolumn{2}{l|}{SelfAtt}
& 96.0 & 93.1 & 75.4 & 88.2 & 20.0 & 95.7 & 92.7 & 75.7 & 88.0 & 7.7 \\
& \multicolumn{2}{l|}{SelfAtt-SyncPE}
& 95.8 & 93.1 & 79.9 & {\bf 89.6} & {\bf 18.7} & 95.6 & 92.9 & 79.5 & {\bf 89.3} & {\bf 6.8} \\
\midrule
\multirow{2}{*}{RWC} & \multicolumn{2}{l|}{SelfAtt}
& 87.5 & 72.0 & 68.8 & 76.1 & 19.2 & 87.5 & 72.2 & 69.1 & 76.3 & 15.5 \\
& \multicolumn{2}{l|}{SelfAtt-SyncPE}
& 87.5 & 76.4 & 72.6 & {\bf 78.8} & {\bf 17.0} & 88.0 & 75.6 & 72.9 & {\bf 78.8} & {\bf 13.2} \\
\bottomrule
\end{tabular}
\label{table:proposed_method_with_PE}
\vspace{-2mm}
\end{table*}

\subsection{Evaluation of Language Modeling}
\label{sec:evaluation_of_language_model}

We evaluated the three language models 
 (bi-gram, GRU, and MLM(-SyncPE) described in Section~\ref{sec:language_models})
 in terms of the perplexities
 for the 203 pieces in the Slakh dataset 
 and the 65 songs in the RWC dataset.
The perplexity for a drum score $\tilde\mY$ is defined as follows:
\begin{align}
\mathrm{PPL}^{*}(\tilde\mY) &= 2^{\frac{1}{N}\mathcal{L}_{\mathrm{lang}}^{*}(\tilde\mY)},
\end{align} 
 where ``*'' denotes ``uni`` or ``bi`` and 
 $\mathcal{L}_{\mathrm{lang}}^{\mathrm{uni}}(\tilde\mY)$
 and $\mathcal{L}_{\mathrm{lang}}^{\mathrm{bi}}(\tilde\mY)$
 are given by Eqs.~(\ref{eq:l_lang_uni}) and (\ref{eq:l_lang_bi}), respectively.
Since $\mathcal{L}_{\mathrm{lang}}^{\mathrm{bi}}(\tilde\mY)$ based on the MLM
 does not exactly give the likelihood for $\tilde\mY$ because of the bidirectional nature,
 unlike $\mathcal{L}_{\mathrm{lang}}^{\mathrm{uni}}(\tilde\mY)$ 
 based on the bi-gram or GRU model with the autoregressive nature,
 $\mathrm{PPL}^{\mathrm{bi}}(\tilde\mY)$ can be only roughly compared
 with $\mathrm{PPL}^{\mathrm{uni}}(\tilde\mY)$.
 
As shown in Table~\ref{table:PPLs},
 the predictive capability of the GRU model 
 was significantly better than that of the bi-gram model
 because the bi-gram model is based on the strong assumption 
 that drum patterns are repeated with the 4/4 time signature.
The MLM using the proposed positional encodings, denoted by MLM-SyncPE,
 slightly outperformed the MLM using the conventional encodings.
The larger the training dataset,
 the lower (better) the perplexity.
The pseudo-perplexity calculated by the MLM was close to 1,
 meaning that the MLM accurately predicted the activations at masked tatums
 from the forward and backward contexts.

\begin{figure}[t]
\centerline{
\includegraphics[width=\linewidth]
{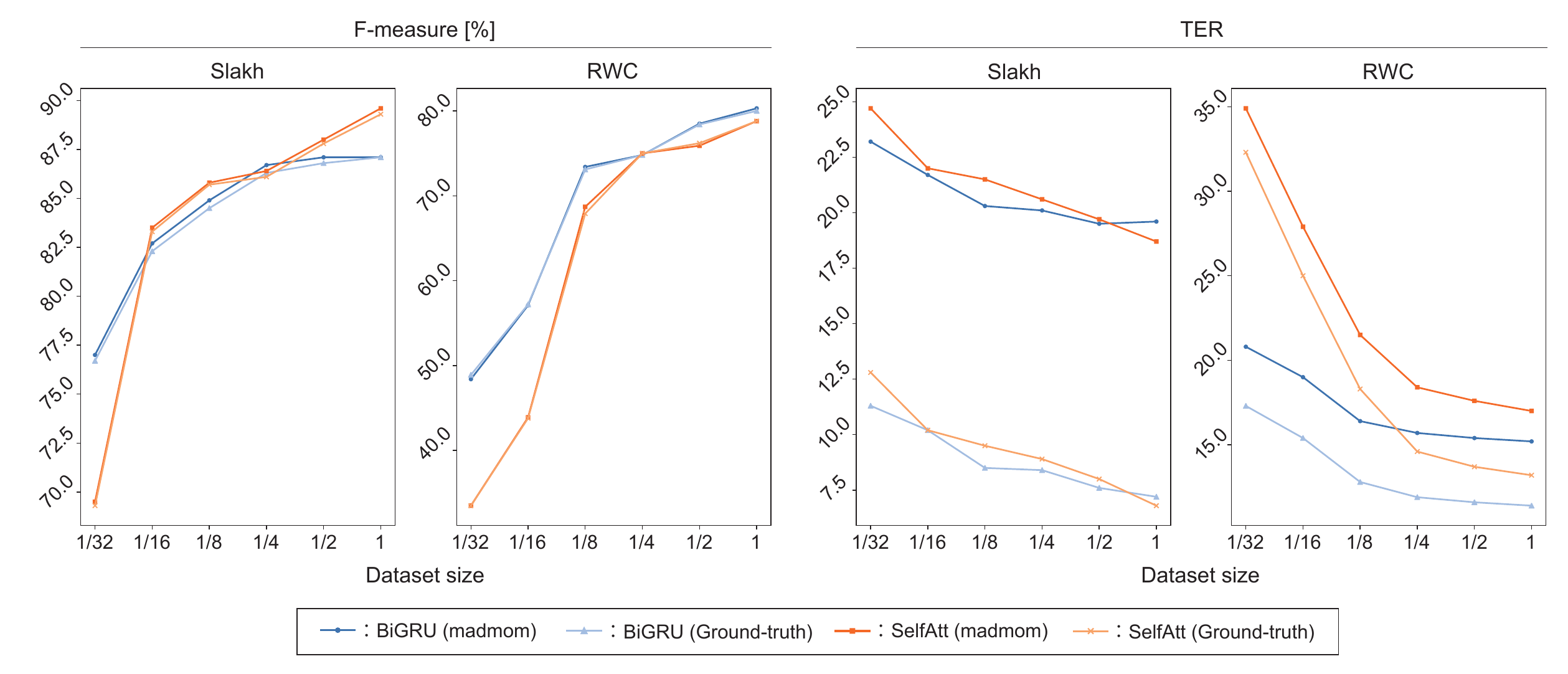}
}
\caption{
F-measures (\%) and TERs obtained by CNN-BiGRU and CNN-SelfAtt-SyncPE
 trained with 1/32, 1/16, 1/4, 1/2, and all of the training data.
}
\label{fig:dataset_size}
\end{figure}

\begin{table*}[t]
\centering
\caption{
F-measures ($\%$) and TERs obtained by CNN-SelfAtt-SyncPE
 regularized by the language models pretrained 
 from 51 or 512 external drum scores.
}
\begin{tabular}{c|ll|ccccc|ccccc} \toprule
& \multicolumn{2}{l|}{Decoder} & \multicolumn{5}{c|}{madmom} & \multicolumn{5}{c}{Ground-truth} \\
& \multicolumn{2}{l|}{~~+ Language model} & BD & SD & HH & Total & TER & BD & SD & HH & Total & TER \\\midrule
\multirow{7}{*}{Slakh} & \multicolumn{2}{l|}{SelfAtt-SyncPE}
& 95.8 & 93.1 & 79.9 & 89.6 & 18.7 & 95.6 & 92.9 & 79.5 & 89.3 & 6.8 \\
&
~~+ Bi-gram & (51)
& 95.1 & 93.0 & 80.5 & 89.5 & 18.6 & 94.9 & 93.1 & 80.6 & 89.5 & 6.5 \\
&
~~+ GRU & (51)
& 96.3 & 92.1 & 80.0 & 89.5 & 18.6 & 96.1 & 92.1 & 79.9 & 89.3 & 6.4 \\
&
~~+ MLM-SyncPE & (51)
& 95.8 & 93.5 & 79.9 & 89.7 & 19.0 & 95.8 & 93.7 & 79.8 & 89.8 & 7.0 \\ 
&
~~+ Bi-gram & (512)
& 95.8 & 92.7 & 80.5 & 89.7 & {\bf 18.5} & 95.4 & 93.1 & 80.5 & 89.7 & {\bf 6.3} \\
&
~~+ GRU & (512)
& 96.4 & 93.0 & 80.5 & {\bf 90.0} & {\bf 18.5} & 96.1 & 92.9 & 80.4 & 89.8 & 6.4 \\
&
~~+ MLM-SyncPE & (512)
& 96.1 & 93.2 & 80.7 & {\bf 90.0} & {\bf 18.5} & 95.8 & 93.3 & 80.9 & {\bf 90.0} & {\bf 6.3} \\ 
\midrule
\multirow{7}{*}{RWC} & \multicolumn{2}{l|}{SelfAtt-SyncPE}
& 87.5 & 76.4 & 72.6 & 78.8 & 17.0 & 88.0 & 75.6 & 72.9 & 78.8 & 13.2 \\
&
~~+ Bi-gram & (51)
& 86.1 & 76.5 & 66.8 & 76.5 & 17.1 & 85.7 & 76.6 & 67.4 & 76.6 & 13.1 \\
&
~~+ GRU & (51)
& 85.6 & 73.3 & 70.2 & 76.4 & 19.6 & 86.7 & 72.9 & 71.3 & 76.9 & 15.7 \\
&
~~+ MLM-SyncPE & (51)
& 86.8 & 75.7 & 71.5 & 78.0 & 17.5 & 86.9 & 75.2 & 72.2 & 78.1 & 13.6 \\ 
&
~~+ Bi-gram & (512)
& 86.0 & 76.5 & 69.6 & 77.4 & 16.8 & 87.0 & 76.0 & 69.6 & 77.5 & 12.7 \\
&
~~+ GRU & (512)
& 87.6 & 76.2 & 73.2 & 79.0 & 16.6 & 87.8 & 76.3 & 73.9 & {\bf 79.4} & 12.5 \\
&
~~+ MLM-SyncPE & (512)
& 88.1 & 74.9 & 75.4 & {\bf 79.5} & {\bf 16.2} & 87.9 & 74.3 & 71.7 & 78.0 & {\bf 12.4} \\ 
\bottomrule
\end{tabular}
\label{table:proposed_method_with_LMs}
\vspace{-2mm}
\end{table*}

\subsection{Evaluation of Drum Transcription}

We evaluated the two transcription models 
 (CNN-BiGRU and CNN-SelfAtt-SyncPE described in Section~\ref{sec:transcription_models})
 that were trained with and without the regularization mechanism
 based on each of the three language models 
 (bi-gram, GRU, and MLM-SyncPE described in Section~\ref{sec:language_models}).
For comparison,
 we tested the conventional frame-level ADT method
 based on a CNN-BiGRU model~\cite{vogl2017drum}
 that had the same architecture 
 as our tatum-level CNN-BiGRU model (Fig.~\ref{fig:encoder_decoder})
 except that the max-pooling layers were not used.
It was trained such that the following frame-level cross entropy was minimized:
\begin{align}
\mathcal{L}_{\mathrm{tran}^{\star}}(\bm\phi^{\star}|\mY^{\star}) 
&=
- \sum_{m=1}^{M} \sum_{t=1}^{T}
\bigl(\beta^{\star}_m Y^{\star}_{m,t}\log \phi^{\star}_{m,t} 
+
(1 - Y^{\star}_{m,t}) \log (1 - \phi^{\star}_{m,t}) \bigr),
\end{align}
where
 $\bm\phi^{\star}, \mY^{\star} \in {\mathbb R}^{M \times T}$
 are the estimated onset probabilities
 and the ground-truth binary activations, respectively,
 and $\beta^{\star} > 0$ is a weighting factor.
For each drum $m$, a frame $t$ was picked as an onset if
\begin{flalign*}
\text{1.\quad}& \phi^{\star}_{m,t} = \max \{ \phi^{\star}_{m, t-w_1:t+w_2}  \}, &\\
\text{2.\quad}& \phi^{\star}_{m,t} \geq \mathrm{mean} \{ \phi^{\star}_{m, t-w_3:t+w_4}  \} + \hat{\delta}, &\\
\text{3.\quad}& t - t_{\mathrm{prev}} > w_5,
\end{flalign*}
where $\hat{\delta}$ was a threshold,
 $w_{1:5}$ were interval parameters,
 and $t_{\mathrm{prev}}$ was the previous onset frame.
These were set to $\hat{\delta} = 0.2$,
 $w_1 = w_3 = w_5 = 2$,
 and $w_2 = w_4 = 0$,
 as in~\cite{vogl2017drum}.
The weighting factors $\beta$, $\gamma$, and $\beta^{\star}$ 
 were optimized
 for the validation data in the Slakh dataset
 and by 10-fold cross validation in the RWC dataset,
 as shown in Table~\ref{table:weighting_factors}.
To measure the tatum-level transcription performance,
 the estimated frame-level onset times
 were quantized at the tatum level
 with reference to the estimated or ground-truth tatum times.
 
As shown in Table~\ref{table:proposed_method_measures},
 CNN-SelfAtt-SyncPE
 worked best for the Slakh dataset
 and CNN-BiGRU with the MLM-SyncPE-based regularization
 worked best for the RWC dataset
 in terms of the F-measure and TER.
This suggests that a sufficient amount of paired data
 are required to draw the full potential of the self-attention mechanism.
CNN-SelfAtt-SyncPE tended to yield higher or comparable F-measures
 for BD and SD in the RWC dataset
 and lower F-measures for HH in both the datasets.
This was because
 percussive sounds in the high frequency range
 were not cleanly separated by the Spleeter
 even with some noises,
 or SelfAtt-based model had a tendency
 to detect other instruments similar to HH such as Maracas.
The MLM-SyncPE-based regularization made a little improvement for the Slakh dataset
 because even the non-regularized model worked well
 on the synthesized dataset with the limited acoustic and timbral variety.
In contrast, it made a significant improvement
 over the bi-gram- or GRU-based regularization for the RWC dataset,
 but 
 required much longer training time
 because every iteration costs $O(C \times N)$,
 where $C$ represents the batch size.
Note that if enough memory is available,
 the MLM-based regularization can be calculated
 in a parallel manner by $O(1)$.
CNN-BiGRU with the MLM-based regularization required the longest training time
 in spite of the highest performance.
The frame-level CNN-BiGRU model~\cite{vogl2017drum}
 required much longer training time
 than our frame-to-tatum model.

\subsection{Investigation of Self-Attention Mechanism}

We further investigated the behaviors of the self-attention mechanisms
 used in the transcription and language models.
To validate the effectiveness 
 of the proposed tatum-synchronous positional encodings (SyncPE),
 we compared two versions of the proposed transcription model,
 denoted by CNN-SelfAtt and CNN-SelfAtt-SyncPE,
 in terms of the F-measure and TER.
As shown in Table~\ref{table:proposed_method_with_PE},
 CNN-SelfAtt-SyncPE always outperformed CNN-SelfAtt by a large margin.
To investigate the impact of the data size 
 used for training the transcription models (CNN-BiGRU and CNN-SelfAtt-SyncPE),
 we compared the performances obtained 
 by using 1/32, 1/16, 1/4, 1/2, and all of the training data in the Slakh or RWC dataset.
As shown in Fig.~\ref{fig:dataset_size},
 CNN-SelfAtt-SyncPE was severely affected by the data size
 and CNN-BiGRU worked better than CNN-SelfAtt
 when only a small amount of paired data were available.
To investigate the impact of the data size 
 used for training the language models (bi-gram, GRU, and MLM-SyncPE),
 we compared the performances obtained by CNN-SelfAtt-SyncPE
 that was regularized with a language model
 pretrained with 512 or 51 external drum scores.
As shown in Table~\ref{table:proposed_method_with_LMs},
 the pretrained language models
 with a larger number of drum scores
 achieved higher performances.
The effect of the MLM-SyncPE-based regularization severely depends on the data size,
 whereas the bi-gram model was scarcely affected by the data size.

\begin{figure}[t]
\centerline{
\includegraphics[width=\linewidth]
{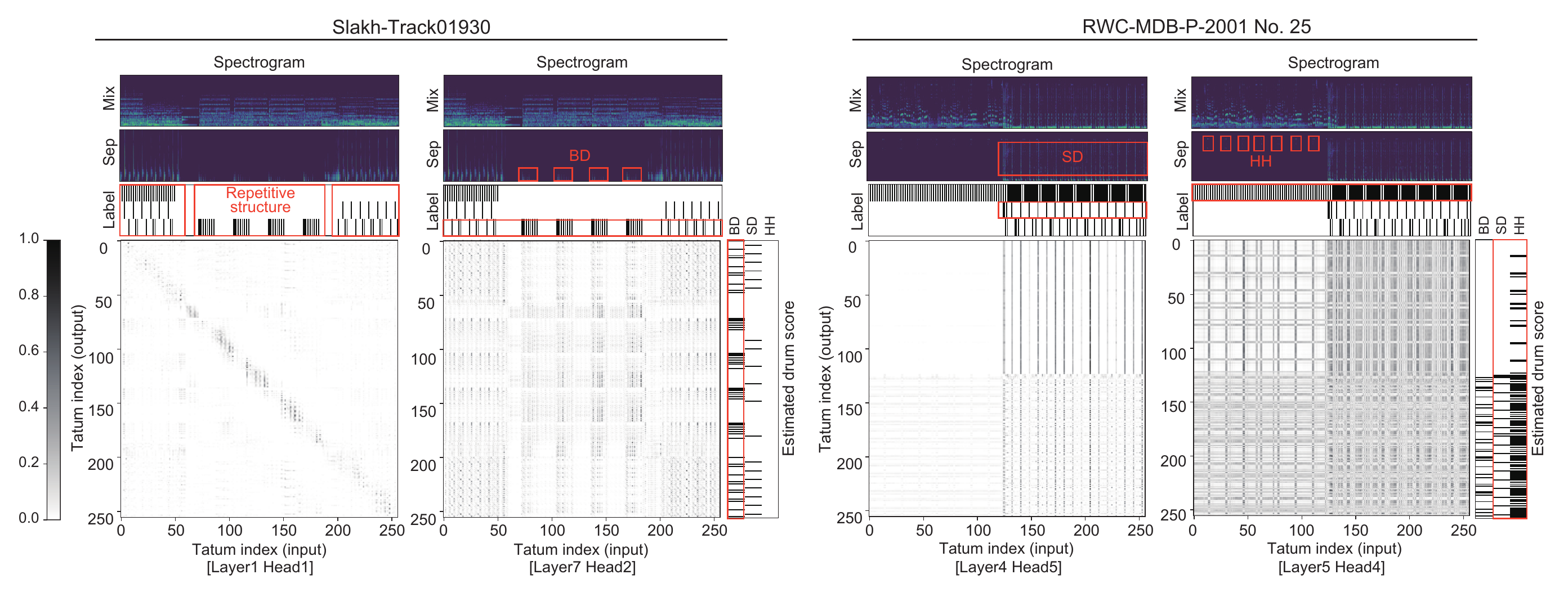}
}
\caption{
Examples of attention matrices representing the repetitive structures of drum scores.
}
\label{fig:attention_matrix}
\end{figure}

\begin{figure}[t]
\centerline{
\includegraphics[width=\linewidth]
{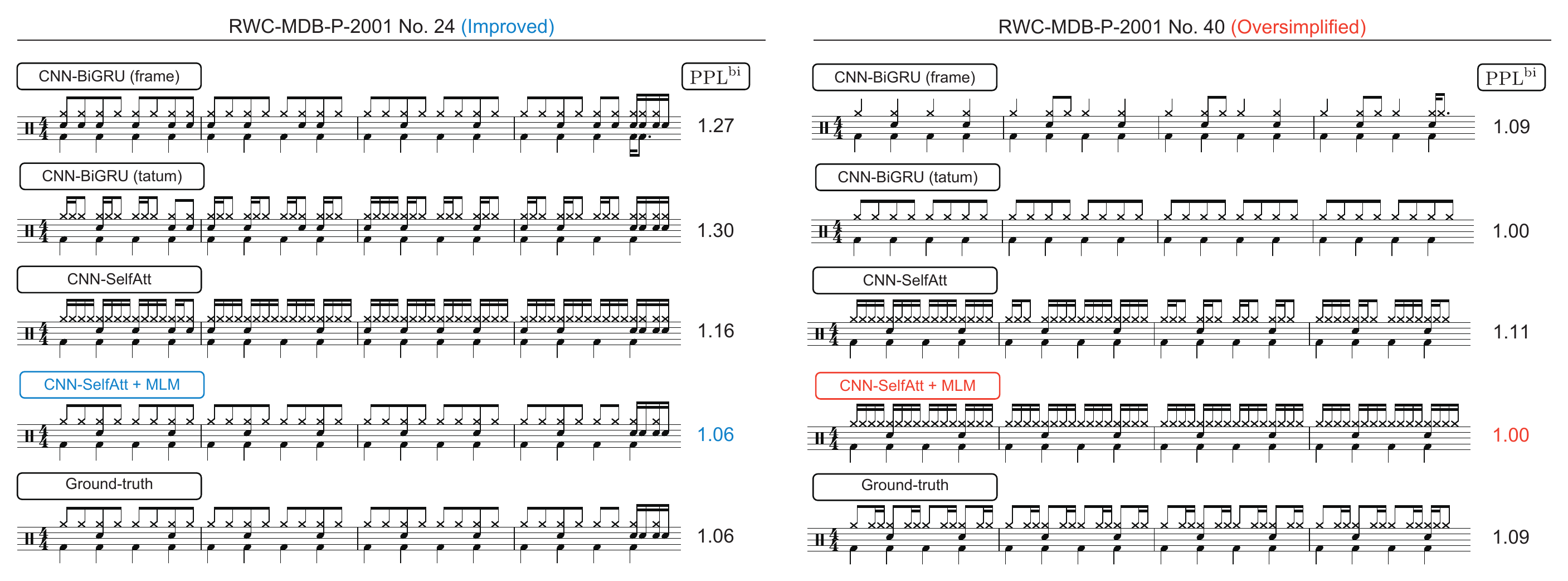}
}
\caption{
Examples of improved (left) and oversimplified (right) drum scores estimated 
 by CNN-SelfAtt-SyncPE with the MLM-SyncPE-based regularization.
}
\label{fig:score_examples}
\end{figure}

We confirmed that CNN-SelfAtt-SyncPE
 with the MLM-SyncPE-based regularization
 learned the global structures of drum scores through attention matrices
 and yielded globally-coherent drum scores.
Fig.~\ref{fig:attention_matrix} shows examples of attention matrices.
In Slakh-Track01930,
 both the global structure of drums
 and the repetitive structure of BD were learned successfully.
In RWC-MDB-P-2001 No. 25,
 the repetitive structures of SD and HH were captured.
These examples demonstrate that the attention matrices at each layer and head
 can capture the different structural characteristics of drums.
Fig.~\ref{fig:score_examples} shows examples of estimated drum scores.  
In RWC-MDB-P-2001 No. 25,
 the MLM-SyncPE-based regularization improved $\mathrm{PPL^{bi}}$ (musical unnaturalness)
 and encouraged CNN-SelfAtt-SyncPE
 to learn the repetitive structures of BD, SD, and HH.
In RWC-MDB-P-2001 No. 40, in contrast,
 although the MLM-SyncPE-based regularization also improved $\mathrm{PPL^{bi}}$,
 it yielded an oversimplified score of HH.

\section{Conclusion}
In this paper, we described a global structure-aware frame-to-tatum ADT method 
 based on self-attention mechanisms.
The transcription model consists of a frame-level convolutional encoder
 for extracting the latent features of music signals
 and a tatum-level self-attention-based decoder
 for considering musically meaningful global structure,
 and is trained in a regularized manner based on an pretrained MLM.
Experimental results showed that
 the proposed regularized model
 outperformed the conventional RNN-based model
 in terms of the tatum-level error rate and the frame-level F-measure,
 even when only a limited amount of paired data were available
 so that the non-regularized model underperformed the RNN-based model.
The attention matrices revealed that
 the self-attention-based model could learn
 the global and repetitive structure of drums
 at each layer and head.
In future work,
 we plan to deal with
 more sophisticated and/or non-regular drum patterns (\eg, fill-ins)
 played using various kinds of percussive instruments (\eg, cymbals and toms). 
Considering that beat and downbeat times are closely related to drum patterns,
 it would be beneficial to integrate beat tracking into ADT
 in a multi-task learning framework. 

\section*{Acknowledgment}
This work was partially supported
 by JST ACCEL No.~JPMJAC1602 and PRESTO No.~JPMJPR20CB
 and JSPS KAKENHI No. 16H01744, No. 19H04137, and No.~20K21813.

\reftitle{References}
\externalbibliography{yes}
\bibliography{refs}

\end{document}